\documentclass[prb,preprint,superscriptaddress]{revtex4-2} 
\usepackage{graphicx} 
\usepackage{amsmath}
\newcommand{\bra}[1]{\langle{#1}|}
\newcommand{\ket}[1]{|{#1}\rangle}

\begin{document}
\title{Factorization with a logarithmic energy spectrum of a central potential}
\author{Ferdinand Gleisberg}
\affiliation{Institut f\"ur Quantenphysik and Center for Integrated Quantum Science and Technology ($\rm {IQ}^{\rm {ST}}$), Universit\"at Ulm, D--89069 Ulm, Germany}
\email{ferdinand.gleisberg@alumni.uni-ulm.de}

\author{Wolfgang P. Schleich}
\affiliation{Institut f\"ur Quantenphysik and Center for Integrated Quantum Science and Technology ($\rm {IQ}^{\rm {ST}}$), Universit\"at Ulm, D--89069 Ulm, Germany}
\affiliation{~Hagler Institute for Advanced Study at Texas A \& M University, Texas A \& M AgriLife Research, 
Institute for Quantum Studies and Engineering (IQSE) and
Department of Physics and Astronomy, Texas A \& M University,
College Station, Texas 77843-4242, USA}

%
%
%

\begin{abstract}
We propose a method to factor numbers based on two interacting bosonic atoms
in a central potential where the single-particle spectrum
depends logarithmically on the radial quantum numbers of the
zero angular momentum states. The bosons
initially prepared in the ground state are excited by a sinusoidally
time-dependent interaction
into a state characterized by the
quantum numbers which represent the factors of a number encoded
in the frequency of the perturbation. We also discuss the full
single-particle spectrum and limitations of our method caused by decoherence.
\end{abstract}



\maketitle



\section{Introduction}\label{Intro}

It is well-known that
the decomposition of a positive integer into a product
of prime factors is a difficult problem 
in number theory  for it requires non-polynomial time 
on a classical computer making it attractive for cryptological
applications.\cite{Hardy} {\it E.g.} for decoding a message encoded by the
famous RSA protocol \cite{RSA} decomposition of a large
semiprime {\it i.e.} a number composed by two primes 
in a reasonable time is needed.
Such a decomposition can easily be prevented by choosing
larger and larger semiprimes. 
If the topic of prime factorization is mentioned somewhere, may be in a discussion or
may be in an article, it does not take long until the name
Peter Shor appears
because on a large ideal
quantum computer Shor's factorizing algorithm \cite{Shor} 
takes only polynomial time
and
is therefore expected to break 
the RSA scheme in the future. 

As an alternative method
we have studied the factorization of integers using 
bosonic atoms in one- and two-dimensional potentials both with a
logarithmic energy spectrum.\cite{Gleisberg2013, Gleisberg2018,Gleisberg2015}
Bosons in a spherically symmetric {\em harmonic}
potential as well as in a spherical box provide
textbook examples  for the study of 
thermodynamics of the Bose--Einstein condensation.\cite{Pethick,Bagnato2019}

Our present theoretical study is motivated by the possibility
to create and control nearly any kind of traps using adiabatic potentials 
as was stated in Ref. \onlinecite{Garraway2016}.
For the presentation of our work we have chosen here
a pedagogical approach.
We constructed numerically a central potential with a logarithmic
energy spectrum.  
Two bosons originally trapped in the ground state of this potential
are excited
by a periodic perturbation with a frequency which contains the semiprime
we want to factor. After some time the bosons are found with a probability
of about one half in a state where
the energies of the individual bosons contain the factors of the
semiprime. Then a measurement of these energies provides
the factors we are looking for. The spherical symmetry of 
the unperturbed potential 
is crucial for our protocol. Among the difficulties to realize
spherical symmetry experimentally we mention that here 
an environment free of gravity 
is required.\cite{Lundblad}

Our article is organized as follows. In Section \ref{Toolbox} we introduce the
logarithmic energy spectrum and discuss the distribution of a given
energy onto two single-particle states.
In Section \ref{1DPot} the Schr{\"o}dinger equation in three dimensions is
solved and it is found that the $s$ states {\it i.e.} the states
with zero azimuthal quantum number are sufficient to determine the
potential with a logarithmic energy spectrum. In Section \ref{Dimensions} we take
into account the boundary condition at the origin and demonstrate that
the single-particle $s$ states exhibit an energy spectrum similar
to the one introduced in Section \ref{Toolbox}. Section \ref{TimeDepPert} discusses the realization
of our factorizing scheme by two bosonic atoms moving in the
central potential determined in Section \ref{1DPot}, moreover, when
excited by a time dependent
interaction the bosons achieve a
transition into the factor state. 
In Section \ref{AppSol} we present the
solution of the
Schr{\"o}dinger equation within the rotating wave approximation
while the calculation can be found in  App. \ref{coupling} and \ref{RWA}.
We mention that after a measurement of the single particle energies
at randomly chosen times the factor state is found with a
probability of about one half. Limitations of our method caused by
decoherence are discussed in Section \ref{Limitations} followed by a short summary.
An elementary discussion  of the absence
of accidental degeneracy in  our logarithic spectrum can be found in App. \ref{accidental}.

\section{Mathematical toolbox}\label{Toolbox}

In the present section we first introduce the logarithmic energy spectrum and
discuss its special role in finding the factors of an integer. 
We then turn
to the distribution of a given energy onto two subsystems.
This discussion constitutes the foundation for our 
factorization protocol.

Our factorization scheme is based on a logarithmic
energy spectrum of the type
\begin{equation}\label{SinglePartSpect}
E_k(L) \equiv \hbar \omega_0 \ln\left(\frac{k}{L}+1\right),\quad k=0,1,2,\ldots
\end{equation}
with $E_0(L)=0$.
Here, the constant $L$ plays the role of a scaling parameter and $\hbar\omega_0$
is the unit of energy. 

In order to find the factors of a given semiprime $N=q_1\cdot q_2$
we try to distribute the energy
\begin{equation}\label{2PartE}
E_{\rm total}(N;L) \equiv\hbar\omega_0\,\ln\left(\frac{N}{L^2}\right)
\end{equation}
onto {\em two} subsystems with spectrum (\ref{SinglePartSpect})
and get
\begin{eqnarray}\label{Distrib}
E_{\rm total}(N;L) & = & \hbar\omega_0\ln\left(\frac{q_1}{L}\right)+\hbar\omega_0\ln\left(\frac{q_2}{L}\right)\\
\label{Distrib2}
&=& E_{q_1-L} + E_{q_2-L}
\end{eqnarray}
where we have used Eq. (\ref{SinglePartSpect}). Since the parameter $L$ 
appears in the indices of the energies in  Eq. (\ref{Distrib2}) it has to be integer.
No negative indices are present in Eq. (\ref{SinglePartSpect})
therefore $N$ must not contain
factors $q_i<L$. Moreover, a factor $q_i=L$ causes the unwanted case that
the total energy (\ref{Distrib}) may be transferred to one
subsystem while the other one is in the ground state $E_0(L)$
and no factorization takes place. We conclude that we have to remove factors
$2,3,\ldots  L$ what can be done by simple division before our
factorization protocol can be applied. However, if $L$ is chosen 
to be unity
it is easily verified that here the trivial
factorization $N=1\times N$ cannot be excluded. 
Moreover, in Section \ref{Dimensions} we shall
see that $L$ has to be odd.
Therefore,
throughout our article we consider the case $L\ge 3$. 
The question of uniqueness of the distribution (\ref{Distrib})
is easily answered
because the fundamental theorem of arithmetics guarantees that the decomposition
of the integer $N$ is unique if both factors, $q_1$ and $q_2$, respectively, 
are prime. 

For our factorization protocol the subsystems have to be brought into
a state with total energy (\ref{Distrib}) followed by a measurement
of the energies of the subsystems which easily allows
the determination of the factors $q_i$ as is described in Sect. \ref{AppSol}.
In the remainder of our article we shall concentrate on the
factorization of semiprimes.

\section{From three dimensions to one dimension}\label{1DPot}

In the present section we realize the subsystem with spectrum (\ref{SinglePartSpect})
by a particle of mass $\mu$ moving in three dimensions in a central potential $V(r)$ 
which we shall determine.

We start with the Schr\"odinger equation in spherical polar coordinates
\begin{equation}\label{SchrEq}
\left[-\frac{\hbar^2}{2\mu}\Delta + V(r)-E\right] \varphi(r,\Theta,\Phi)=0
\end{equation}
and consider the wave functions 
\begin{equation}\label{3DFunction}
\varphi_{k,\ell,m}(r,\Theta,\Phi)\equiv R_{k,\ell}(r)\,Y_\ell^m(\Theta,\Phi)
\end{equation}
which are simultaneous
eigenfunctions of the Hamiltonian $\hat H$, the square of the angular momentum $\hat L^2$, and its
$z$-component $\hat L_z$ which form a complete commuting set of operators
with eigenvalues $E_{k,\ell}$, $\hbar^2\,\ell(\ell+1)$ and $\hbar\, m$, respectively. The
radial quantum number $k$ as well as the 
azimuthal quantum number $\ell$ takes values $0,1,2,\ldots$ while the magnetic quantum number $m$
takes the $2\ell+1$ values $-\ell\ldots\ell$. The
functions $Y_\ell^m(\Theta, \Phi)$ are the spherical harmonics.
In what follows we shall use a short-hand notation for the three quantum numbers
${\bf k}\equiv (k,\ell,m)$.

Because the solution of Eq. (\ref{SchrEq}) can be found in most textbooks we jump directly
to the radial equation valid in the region $r \ge  0$
\begin{equation}\label{RadialEq}
\left[-\frac{\hbar^2}{2\mu}\frac{1}{r}\frac{d^2}{dr^2}r+\frac{\hbar^2\,\ell(\ell+1)}{2\mu r^2}+V(r)-E_{k,\ell}\right]\,R_{k,\ell}(r)=0
\end{equation}
with the condition that $R_{k,\ell}(r)$ has to be square integrable and finite at the origin $r=0$.
We consider $s$ states ($\ell=0$) and set 
\begin{equation}\label{RadialFkt}
R_{k,0}(r)=\frac{u_{k,0}(r)}{r} 
\end{equation}
with the boundary condition
\begin{equation}\label{bc}
u_{k,0}(0)=0.
\end{equation}
Moreover, we write the cartesian coordinate $x$
for the variable $r$ and assume a symmetric potential $V(x)=V(-x)$ where 
now $-\infty<x<\infty$. With these modifications it is
easy to change Eq. (\ref{RadialEq}) into the equation
\begin{equation}\label{1DSchrEq}
\left[-\frac{\hbar^2}{2\mu}\frac{d^2}{dx^2}+V(x;L)-E_{k}(L)\right]u_{k}(x;L)=0
\end{equation}
where for the moment we do not take into account the boundary conditions (\ref{bc})
and omit the $s$ state index $\ell=0$.
This is the well-known 
Schr\"odinger equation for a particle of mass $\mu$ moving 
in the one-dimensional potential $V(x;L)$ with wave functions $u_{k}(x;L)$
which are even (odd) for even (odd) indices $k$.
Here we have changed our notation in order to
emphasize that the energies $E_{k}(L)$ (\ref{SinglePartSpect}), the potential $V(x;L)$,
and the wave functions $u_{k}(x;L)$ depend on the scaling parameter $L$.
Our iteration algorithm to determine the potential $V(x;L)$ from the single
particle spectrum (\ref{SinglePartSpect}) is based on the Hellmann-Feynman
theorem and is described in a previous article.\cite{Mack2010} 
In Fig. 1 we show $V(x;L=3)$ together with the eigenfunctions $u_k(x;L)$ 
for $0\le k\le 6$. 

\begin{figure}[htb]
\includegraphics[height=8cm,width=8cm]{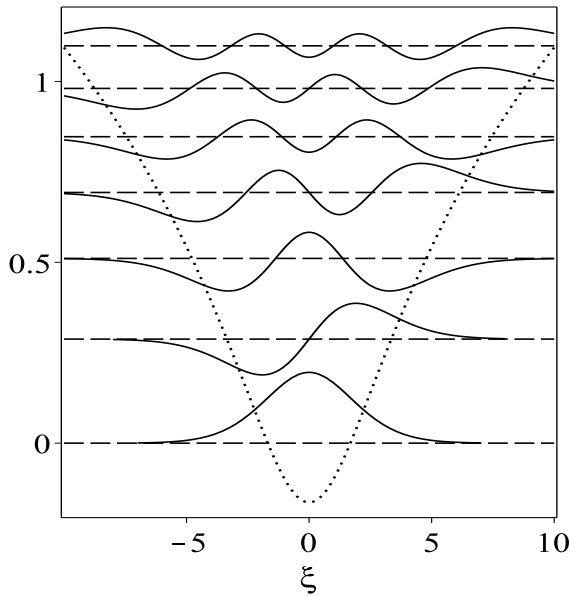}
\caption{\label{Fc1}
One-dimensional potential 
$V(\xi;L)$ (dotted line) 
creating a logarithmic 
energy spectrum 
for a scaling parameter $L=3$ as a function of dimensionless coordinates
$\xi\equiv \alpha \, x$ with $\alpha^2\equiv\mu \omega_0 /\hbar$.
This potential is determined numerically by an iteration algorithm based on a 
perturbation theory using the Hellmann-Feynman theorem 
and is designed to obtain a logarithmic dependence of the energy 
eigenvalues $E_k (L)$ on the quantum number 
$k$ as given in Eq. (\ref{SinglePartSpect}).
In the neighborhood of the origin the potential is approximately
harmonic whereas for large values of $\xi$ it is logarithmic. 
In solid lines we depict the numerically determined 
energy wave functions of the first 7 states in their 
dependence on the dimensionless position. Both, the energies $E_k(L)$, $k=0,1,\ldots 6$ 
(dashed lines) as well as the potential $V(\xi;L)$ 
are shown in units
of $\hbar\omega_0$.
}
\end{figure}

Note that states with quantum numbers $\ell>0$ were not needed
for the determination of the potential $V(x;L)$. Some aspects of the
full spectrum $E_{k,\ell}(L)$, however,
are discussed in  App. \ref{accidental}.

\section{Energy spectrum of $s$ states}\label{Dimensions}

We continue to limit ourselves to $s$ states only and suppress
the index $\ell=0$.
In the last section the potential $V(x,L)$ and the functions $u_k(x,L)$ 
were determined numerically and displayed in Fig. \ref{Fc1}.
The three-dimensional potential $V(r;L)$ as well as the
eigenfunctions $u_k(r;L)$ follow simply by replacing the coordinate $x$
by $r$ in either of them, where now only the region $r\ge 0$ is considered. 
Figure 2 shows the potential $V({\bf r};L)$ with
position vector $\bf r$ in the $x$-$y$ plane. 
\begin{figure}[htb]
\includegraphics[height=8cm,width=8cm]{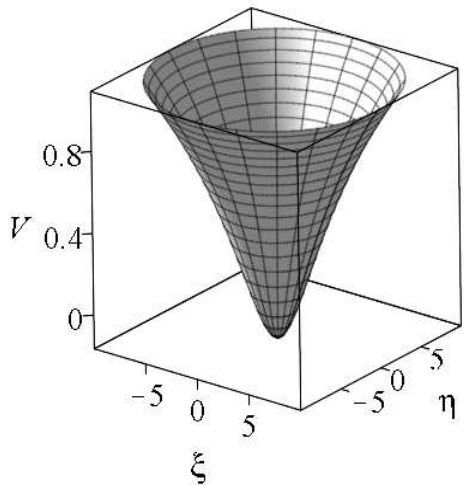}
\caption{\label{Fc2}
Three-dimensional potential $V(r;L=3)$ in units of $\hbar \omega_0$ creating the
logarithmic energy spectrum Eq. (\ref{Etilde}) with scaling parameter
$K=2$ as a function of the
dimensionless coordinates $\xi=\alpha x$ and $\eta=\alpha y$ plotted in
the plane $z=0$.
}
\end{figure}
Here in three dimensions only {\em odd} solutions $u_k(x;L)$
of Eq. (\ref{1DSchrEq}) can satisfy the boundary condition
(\ref{bc}). Therefore, energies $E_{k}(L)$ as well as eigenfunctions $u_k(x;L)$ with
{\em even} index $k$ which were present in one dimension in Eq. (\ref{1DSchrEq}) do not appear
anymore in three dimensions. 

We will show now that the remaining spectrum $E_{k}(L)$ has 
indeed the form of Eq. (\ref{SinglePartSpect}) and therefore guarantees the validity 
of the results of Section \ref{Toolbox} which we need for our factorization
procedure. We rewrite the energies with odd 
radial quantum numbers $k=2j+1$ 
\begin{equation}\label{Rewrite}
E_{2j+1}(L)=\hbar\omega_0 \ln\left(\frac{2j+1}{L}+1\right)
\end{equation}
with $j=0,1,2,3\ldots$ and shift them by $-\hbar\omega_0\ln(1/L+1)$.
It is easy to verify that the new spectrum  is identical with
single particle spectrum
(\ref{SinglePartSpect})
\begin{equation}\label{Etilde}
 E_j^{\rm 3d}(K)=\hbar\omega_0\ln\left(\frac{j}{K}+1\right)
\end{equation}
except that $L$ has to be replaced by a new scaling parameter
\begin{equation}\label{Lprime}
K=\frac{L+1}{2}.
\end{equation}

In order that $K$ is a positive
integer
the parameter $L$ has to be odd. All the statements made in
Section \ref{Toolbox} referring to the scaling length $L$
remain valid here provided $L$ is replaced by $K$.
The eigenfunctions $v_j(r;K)$ belonging to $E_j^{\rm 3d}(K)$
are
\begin{equation}
v_{j}(r;K) \equiv u_{2j+1}(r;L).
\end{equation}
Figure 3 shows 
the radial functions 
\begin{equation}\label{Rad3d}
R_j(r)=\frac{v_j(r;K)}{r}
\end{equation} 
for indices $k=0,\ldots 5$ 
together with the potential $V(r;L=3)$ and the energy levels $E_j^{\rm 3d}(K=2)$ (\ref{Etilde}).
\begin{figure}[htb]
\includegraphics[height=8cm,width=8cm]{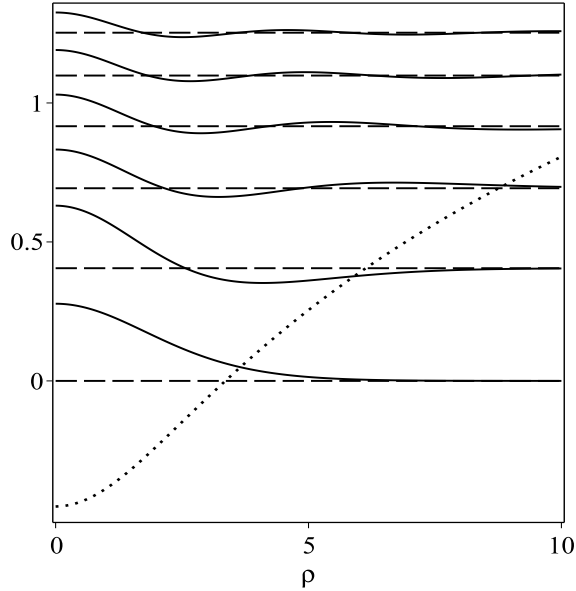}
\caption{\label{Fc3}
Central potential 
$V(\rho;L=3)$ 
creating the loga\-rith\-mic 
energy spectrum $E^{\rm 3d}_j(K=2)$ (\ref{Etilde}) 
in units of $\hbar\omega_0$
as function of the dimensionless radius
$\rho\equiv \alpha \, r$ 
together with the corresponding radial
functions $R_j(\rho)$ (\ref{Rad3d}) of the first 6 states in their 
dependence on the dimensionless radius.  
Note that the energies have been shifted in order that the ground
state has an energy zero here.
}
\end{figure}
To simplify the notation we pass over to the bra-ket formalism.
The single-particle Schr\"odinger equation for the $s$ states reads
\begin{equation}
{\hat H}(K)\, \ket{j} = E_j^{\rm 3d}(K) \,\ket{j}\qquad j=0,1,2,\ldots 
\end{equation}
where the quantum numbers $\ell=m=0$ are suppressed.
The hamiltonian ${\hat H}(K)$ is
characterized by the parameter $K$ (\ref{Lprime}).

The Schr{\"o}dinger equation for two non-interacting bosons is
\begin{equation}
\left({\hat H}_{1,2}(K)-E_{m,n}(K)\right)\, \ket{m,n}_B = 0
\end{equation}
with
\begin{eqnarray}\label{H12}
{\hat H}_{1,2}(K)& = &{\hat H}_1(K)+{\hat H}_2(K)\\
E_{m,n}(K) & =  & E_{m}^{\rm 3d}(K)+ E_{n}^{\rm 3d}(K)\label{Epq}
\end{eqnarray}
in accordance with Eqs. (\ref{Etilde} and \ref{Lprime}). 
Note that bosonic two-particle states are
defined by
\begin{eqnarray}\label{Bosons}
\ket{m,n}_B \equiv \frac{1}{\sqrt{2}}\left(\ket{m,n}+\ket{n,m}\right),\\
\ket{m,m}_B \equiv \ket{m,m}.\nonumber
\end{eqnarray}
If two identical non-interacting bosons are
in a state with energy 
\begin{equation}\label{FactorState}
\hbar\omega_0\,\ln\left(\frac{N}{K^2}\right)=E_{p-K}^{\rm 3d}+E_{q-K}^{\rm 3d} 
\end{equation}
where $N\equiv p\cdot q$ is semi-prime
then according to Eqs. (\ref{Distrib2} and \ref{Etilde})
the bosons are in the state $\ket{p-K,q-K}_B$ we call factor state. 
A measurement of the energy of one of the bosons can only result
in $\hbar \omega_0 \ln(p/K)$ or $\hbar \omega_0 \ln(q/K)$ and immediately
yields the prime factors $p$ and $q$, respectively.
In the remainder of the article
we suppress the scaling parameter $K$ as well as the suffix $B$ and the 
superscript $\rm{3d}$ in order to simplify the notation.

\section{Time dependent perturbation}\label{TimeDepPert}
In the present section we describe how the factorization protocol of
Section \ref{Toolbox} can be realized with two interacting 
identical bosons placed in a three-dimensional potential shown in Fig. 2 with
single particle 
spectrum (\ref{Etilde}) which does not show degeneracy
for $s$ states.
We prepare 
the two bosons in the ground state $\ket{\bf{0,0}}$ and at $t=0$ a perturbation
\begin{equation}\label{Perturb}
\delta V({\bf r}_1,{\bf r}_2;t) = \gamma \sin(\omega_{\rm ext}t) \,w({\bf r}_1,{\bf r}_2)
\end{equation}
is switched on. The frequency $\omega_{\rm ext}$ is chosen later
in a way suitable for the factorization procedure.

The movement of the two-particle ket $\ket{\Psi (t)}$
is now governed by the Schr\"odinger equation
in three dimensions 
\begin{equation}\label{SchrEqPert}
i\hbar\frac{d}{dt} \ket{\Psi(t)}=[{\hat H}_{1,2} +\delta V(t) ]\ket{\Psi(t)}
\end{equation}
with the unperturbed stationary equations
\begin{equation}
{\hat H}_{1,2} \,\,\ket{{\bf k}_1,{\bf k}_2} = E_{{\bf k}_1,{\bf k}_2} \ket{{\bf k}_1,{\bf k}_2}.
\end{equation}
We substitute the expansion of the solution $\ket{\Psi(t)}$ into the
two-particle eigenkets $\ket{{\bf k}_1,{\bf k}_2}$ of the unperturbed Hamiltonian 
${\hat H}_{1,2}$ 
\begin{equation}\label{Expand}
\ket{\Psi(t)}=\sum_{{\bf k}_1,{\bf k}_2} \, b_{\,{\bf k}_1,{\bf k}_2}(t) e^{-i E_{{\bf k}_1,{\bf k}_2} t/\hbar} \, \ket{{\bf k}_1,{\bf k}_2}
\end{equation}
into Eq. (\ref{SchrEqPert})  and arrive at the coupled system
\begin{widetext}
\begin{eqnarray}\label{CoupSys}
\hspace{-1.5cm}
i\hbar\, {\dot b}_{\,{\bf k}_1, {\bf k}_2}(t)=
\gamma \sin(\omega_{\rm ext} t) \sum_{{\bf k}'_1,{\bf k}'_2}
e^{i(E_{{\bf k}_1, {\bf k}_2}-E_{{\bf k}'_1,{\bf  k}'_2}) t/\hbar}\,
  W_{{\bf k}_1 ,{\bf k}_2; {\bf k}'_1, {\bf k}'_2}\, b_{\,{\bf k}'_1,{\bf k}'_2}(t)\\
b_{\,{\bf k}_1,{\bf k}_2}(0)=1  \mbox{ { } for { }  }  k_1+k_2+\ell_1+\ell_2=0 \mbox{ { } and { } } 
b_{\,{\bf k}_1,{\bf k}_2}(0)=0 \mbox{ { } otherwise}\nonumber
\end{eqnarray}
\end{widetext}
which has to be solved for the probability amplitudes $b_{\,{\bf k}_1, {\bf k}_2}(t)$.
Here the indices ${\bf k}_1$ {\it etc.} represent the triple of quantum numbers: 
${\bf k}_1\equiv (k_1,\ell_1,m_1)$ introduced in Section \ref{1DPot}. The eigenkets 
$\ket{{\bf{k}_1},{\bf{k}_2}}$ of ${\hat H}_{1,2}$,
the amplitudes $b_{\,{\bf k}_1,{\bf k}_2}(t)$, and the matrix elements 
\begin{equation}\label{MElem}
W_{{\bf k}_1,{\bf k}_2;{\bf k}'_1,{\bf k}'_2}\equiv 
\bra{{\bf k}_1,{\bf k}_2} w({\bf {\hat r}}_1,{ \bf {\hat r}}_2) \ket{{\bf k}'_1,{\bf k}'_2}
\end{equation}
are 'bosonic' ones
in the sense of Eq. (\ref{Bosons})
and are built out of the 
eigenkets $\ket{{\bf k}_1,{\bf k}_2}$ of ${\hat H}_{1,2}$ and the spacial part $w$ of the
perturbation $\delta {\hat V}$.
Moreover, in the summation in Eqs. (\ref{Expand}) and (\ref{CoupSys}) 
the same states must not be counted twice.

In App. \ref{coupling} we study the matrix element (\ref{MElem})
for the case of a contact interaction between the particles
while in App. \ref{RWA} we benefit from the rotating wave approximation and
reduce the system (\ref{CoupSys}) to the
much simpler system  
(\ref{b00Short}) and (\ref{bpqShort}) of  only two differential
equations for two probability amplitudes, namely that of the ground state
and that of the factor state, respectively.
We solve them in then next section.

\section{Approximate solution}\label{AppSol}

With the help of the so-called secular or rotating wave approximation (RWA) \cite{Cohen}
we have reduced the infinite system (\ref{CoupSys}) to only two first-order differential equations
with constant coefficients (\ref{b00Short} and \ref{bpqShort}).
App. \ref{RWA} presents the calculation. Together with the initial conditions
$b_{0,0}(0)=1$ and $b_{p-K,q-K}(0)=0$ the equations are immediately solved.
The resulting probability amplitudes are
for the ground state
\begin{equation} 
b_{0,0}(t)=\cos(\Omega t)
\end{equation}
 and 
\begin{equation}\label{b_sin}
b_{p-K,q-K}(t)=\sin(\Omega t)
\end{equation}
 for the factor state, respectively, and the
so-called Rabi frequency is
\begin{equation}\label{Rabi}
\Omega=\frac{\gamma}{2\hbar}\,W_{0,0;p-K,q-K}
\end{equation}
which is proportional to the interaction matrix element (\ref{W02}).
In Sect. \ref{Toolbox} and \ref{Dimensions}
it was shown that if the bosons 
are in the factor state $\ket{p-K,q-K}$
they have a
two-particle energy $\hbar\omega_0\ln(N/K^2)$ with $N=p\cdot q$
(\ref{FactorState}).

As mentioned there the factors $p$ or $q$ are determined by
a measurement of single-particle energies (\ref{Distrib})
and the factorization protocol has ended successfully.

At time $t$ the system can be found with probability $|b_{p-K,q-K}(t)|^2$
in the factor state and at times equal to an odd multiple of $\pi / (2\Omega)$
with certainty but, unfortunately, the Rabi frequency $\Omega$ is not known.
Instead, we content ourselves with measuring at a time chosen
at random from a time interval $[0,T]$ much larger than $\pi/\Omega$. Utilizing Eq. (\ref{b_sin}) 
it is easy to see that the probability to find the factor
state is about one half.  Then the measurement of a single-particle energy gives one of the factors
while the other one follows from division.

An estimate for a time of measurement
by making a guess for the factors $p$ and $q$ and determining so the Rabi frequency (\ref{Rabi})
was presented in a previous article.\cite{Gleisberg2013}

\section{Limitations}\label{Limitations}

In the present section we shall sketch what prevents our protocol to
factor larger and larger semiprimes. According to Ref. \onlinecite{Schleich}
 there is a
high probability for the periodic transition into the factor state
as long as the difference between the energies of this state and of the next off-resonant
state, respectively, is larger than the energy $\hbar\Omega$ of the Rabi oscillation
\begin{equation}
\hbar \omega_0\left|\ln\left(\frac{N\pm 1}{K^2}\right)-\ln\left(\frac{N}{K^2}\right)\right|
\approx \frac{\hbar \omega_0}{N} \gg \hbar\Omega .
\end{equation}
Because the Rabi frequency $\Omega$ defined by Eq. (\ref{Rabi}) 
is proportional to the strength
$\gamma$ of the perturbation (\ref{Perturb}) this condition can
easily be satisfied by choosing $\gamma$ as small as needed. Unfortunately, a second condition arises
from Section \ref{AppSol} where the time of measurement of the energies of the
two bosons was chosen randomly from an interval $[0,T]$. To find the factor state
with a probability of $\approx 1/2$ the length $T$ of the interval had to fulfill
the condition $\Omega \,T\gg 1$. On the other hand the system 
has to be {\em free of  decoherence} during the time interval {\it i.e.} $T<T_{\rm dec}$ 
leading to two inequalities the Rabi frequency has to fulfill
\begin{equation}
\Omega \gg \frac{1}{T_{\rm dec}} \quad \mbox{and} \quad \Omega \ll \frac{\omega_0}{N}.
\end{equation}
Our aim is to find an upper limit of the number to
be factored $N$. In our articles Ref.
\onlinecite{Gleisberg2013} and \onlinecite{Gleisberg2015}
for different experimental situations
and models for the spacial part of the interaction 
an $N$-dependence of the transition matrix element
\begin{equation}
W_{0,0;p-K,q-K} \propto N^{-1/2}
\end{equation}
was found in rough approximation and the same is valid, of course,
for the Rabi frequency $\Omega$ (\ref{Rabi}). The semiprime $N$ to be factored
therefore has an upper limit
\begin{equation}
N < \min \left( \left[\frac{\gamma T_{\rm dec}}{\hbar}\right]^2 ,\left[\frac{\hbar\omega_0}{\gamma}\right]^2\right).
\end{equation}
Assuming that according to Eq. (\ref{Perturb}) the interaction strength 
$\gamma$  can be chosen at will 
this relation shows that the crucial limiting factor for the magnitude of $N$
is the decoherence time $T_{\rm dec}$.

\section{Summary}

In the present article we have proposed a method to find the factors of a semiprime $N$
based on the quantum dynamics of two identical bosonic atoms
moving in a spherically symmetric trap whose $s$ states exhibit 
a logarithmic single particle
spectrum. 

In the first part of our work we have determined  a central potential such that it
has a logarithmic energy spectrum. First  we calculated numerically 
a one-dimensional potential from a logarithmic single particle spectrum.
Because of the close relationship bet\-ween three-dimensional spherically
symmetric and one-dimensional problems, respectively, the central
potential then was easily found. As expected  this potential 
had an energy spectrum with a logarithmic $s$ wave part but
with a scaling length different from the one in the one-dimensional
spectrum.

In the second part of our work we attacked the problem how to bring the
bosons into the factor state.
The bosons were excited from their ground state
by a periodic time-dependent contact interaction
of a frequency which was determined by the number $N$ to be factored.
To exclude transitions bet\-ween non-$s$ states we discussed {\it in extenso} the absence
of degeneracy.  Then we showed within the framework of  the 
well-known rotating wave
approximation that the bosons performed a Rabi oscillation
between the ground state and the factor state.
The latter was found with a probability of about one half when
the energies of the bosons were measured at a randomly chosen time.
From these the factors of $N$ were easily determined
and our factorization protocol has ended successfully.

\begin{acknowledgments}
We thank  M. A. Efremov and M. Freyberger for stimulating discussions on this topic.
WPS is grateful to the Hagler Institute for Advanced Study at Texas A\&M University for a
Faculty Fellowship and to Texas A\&M University AgriLife Research for its support.
The research of the IQST is financially supported by the Ministry of 
Science, Research and Arts Baden-W{\"u}rttemberg.
\end{acknowledgments}

\appendix 

\section{Matrix elements of the interaction}\label{coupling}
In this appendix we study the matrix element (\ref{MElem})
\begin{equation}\label{A1}
W_{{\bf k}_1,{\bf k}_2;{\bf k}'_1,{\bf k}'_2} \equiv  \bra{{\bf k}_1,{\bf k}_2} w({\bf {\hat r}}_1,{\bf {\hat r}}_2)
\ket{{\bf k}'_1,{\bf k}'_2}
\end{equation}
assuming a contact interaction
between the particles
\begin{equation}\label{SimpleEx}
w({\bf r}_1,{\bf r}_2) = \delta^{(3)}({\bf r}_1-{\bf r}_2).
\end{equation}
With the help of  (\ref{SimpleEx}) 
the transition matrix element can be represented by the eigenfunctions $\varphi_{\bf k}(\bf r)$ (\ref{3DFunction})
introduced in section \ref{1DPot}:
\begin{equation}\label{A3}
W_{{\bf k}_1,{\bf k}_2;{\bf k}'_1,{\bf k}'_2} \equiv  
\int d^3 r \, \varphi_{{\bf k}_1}({\bf r})^\ast \varphi_{{\bf k}_2}({\bf r})^\ast\, 
\varphi_{{\bf k}'_1}({\bf r}) \varphi_{{\bf k}'_2}({\bf r}).
\end{equation} 
Having in mind that we start our procedure at time $t=0$ with the two particles in the ground state $\ket{{\bf 0},{\bf 0}}$
we consider the matrix elements    $W_{\bf{0,0};\bf{k}_1,\bf{k}_2}$ for a transition  into
some excited state  $\ket{{\bf k}_1,{\bf k}_2}$. It is not difficult to derive the expression 
\begin{widetext}
\begin{equation}\label{W02}
W_{\bf{0,0};\bf{k}_1,\bf{k}_2}=\frac{1}{4\pi}\int\,dr\,r^2\,R_{0,0}(r)^2\,R_{k_1,\ell_1}(r)\,R_{k_2,\ell_2}(r)
\delta_{\ell_1,\ell_2}
\,\delta_{m_1+m_2,0}.
\end{equation}
\end{widetext}

Here we have substituted Eq. (\ref{3DFunction}) for the eigenfunctions $\varphi_{{\bf k}}({\bf r})$,
moreover we applied
the well-known orthonormality of the spherical harmonics
\begin{equation}\label{orthrel}
\int\!\int d\Omega\,Y_{\ell_1}^{m_1\ast}(\theta,\varphi)\,Y_{\ell_2}^{m_2}(\theta,\varphi)=\delta_{\ell_1,\ell_2}\,\delta_{m_1,m_2}
\end{equation}
and the relation for their complex conjugate
\begin{equation}
Y_\ell^{m\ast}(\theta,\varphi)=Y_\ell^{-m}(\theta,\varphi).
\end{equation}
Note that $Y_0^0 \equiv 1/\sqrt{4\pi}$.

In the next appendix we use the matrix element (\ref{W02}) when we return to
the system of coupled equations (\ref{CoupSys}) which we shall simplify considerably.

\section{Rotating wave approximation (RWA)}\label{RWA}

In this appendix we put all magnetic quantum numbers $m_i=0$ and omit them henceforth. 
This assumption 
will be justified in the calculation below. A single-particle state
is now characterized by only two quantum numbers $k$ and $\ell$, respectively.
We study the sub-system
\begin{eqnarray}\label{GenSys}
i\hbar\, {\dot b}_{\,0,0;0,0}(t) & = &
\gamma \sin(\omega_{\rm ext} t) \\
 & & \times \sum_{k_1,k_2,\ell}
e^{-i(E_{k_1,\ell}+E_{k_2,\ell}) t/\hbar}  \nonumber \\
 & &  \times \,\,\, W_{0,0;0,0;k_1,\ell,k_2,\ell} \,\, b_{\, {k_1},\ell;k_2,\ell}(t) \nonumber
\end{eqnarray}
of the system (\ref{CoupSys}) with the matrix element (\ref{W02}) and a zero
ground state energy of the two bosons. 

The essence of the RWA applied to (\ref{GenSys}) is simply
to keep all terms with constant coefficients on the right hand side
and to neglect all oscillating terms. 
The external frequency $\omega_{\rm ext}$ is now chosen such
that the energy $\hbar\omega_{\rm ext}$ agrees with the energy 
\begin{equation}\label{omegaExt}
E_{p-K,0;q-k,0}=E_{p-K,0}+E_{q-K,0}=\hbar\omega_0\ln\left(\frac{N}{K^2}\right)
\end{equation}
of the factor state and is determined by the number to be factored $N=p\cdot q$.
Consider now the time dependent factors 
\begin{eqnarray}\label{constant}
& & \hspace{-0.4cm} \frac{1}{2i}\,\left[e^{i(E_{p-K,0}+E_{q-K,0})t/\hbar}-e^{-i(E_{p-K,0}+E_{q-K,0})t/\hbar}\right] \\
& &  \times \,\, e^{-i(E_{k_1,\ell}+E_{k_2,\ell})t/\hbar} \nonumber
\end{eqnarray}
which appear on the right hand side of (\ref{GenSys}). Note here the expanded sinus.
Assuming $p\ge q$ only the term with $k_1=p-K$, $k_2=q-K$ and $\ell=0$ survives
the application of the RWA and is 
of amount $(2i)^{-1}$. 
Appendix \ref{accidental} discusses the absence of accidental degeneracy
in the single particle spectrum $E_{k,\ell}$ (\ref{SinglePartSpect}) which is demonstrated in
Figure \ref{Fc7}. 
None of the terms with $\ell\ge 1$ may therefore lead to additional constant terms in (\ref{constant}).
We note in passing that the $(2\ell+1)$-fold degeneracy with respect to the magnetic
quantum number $m$ is simply unity as was mentioned above.

With these results it is easy to see  that (\ref{GenSys}) is reduced to the equation
\begin{equation}\label{b00Short}
i\hbar{\dot b}_{\,0,0}(t)  =
 \frac{\gamma}{2i}  W_{0, 0 ; {p-K}, {q-K} }\, b_{ p-K, q-K}(t)
\end{equation}
where the index $\ell=0$ present in the matrix elements and in the probability 
amplitudes is omitted here for convenience.
To derive a second equation 
we select  the term with $k_1=p-K$ and $k_2=q-K$
from (\ref{CoupSys}) and proceeding like before
we get
\begin{equation}\label{bpqShort}
i\hbar{\dot b}_{\,p-K,q-K}(t)  =
 -\frac{\gamma}{2i}  W_{p-K, q-K ; 0,0 }\, b_{ 0, 0}(t).
\end{equation}
of the unperturbed $s$ states. Equations (\ref{b00Short} and \,\ref{bpqShort})  characterize the dynamics of the
two-boson system driven by the periodic perturbation (\ref{Perturb}) 
with frequency (\ref{omegaExt}). 
Together with the initial conditions $b_{0,0}(0)=1$ and $b_{p-K,q-K}(0)=0$ 
and the symmetry
\begin{equation}\label{Wsymm}
W_{m,n;0,0}=W_{0,0;m,n}
\end{equation}
they are solved
in Sect. \ref{AppSol}.
\begin{figure}[htb]
\includegraphics[height=8.0cm,width=8.0cm]{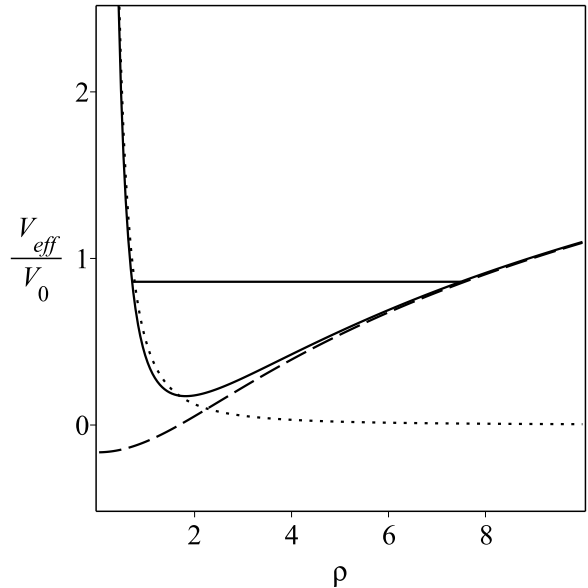}
\caption{\label{Fc4}
Scaled effective potential (solid line) formed by the
angular momentum barrier (dotted)
and the potential
$V(\rho;L=3)$ (dashed)
which in the quantum case creates the loga\-rith\-mic 
energy spectrum (\ref{Etilde}) 
for a scaling parameter $K=2$ (\ref{Lprime}) as function of the dimensionless radius
$\rho\equiv \alpha_{\rm cl} \, r$. 
The horizontal line $E=0.86\, V_0$ denotes the energy 
of the radial coordinate $r(t)$ of the classical particle moving {\em periodically} from the left
turning point to the right one and back. Of course, $\Theta(t)$ is {\em not} 
periodic as is the orbit $r(\Theta)$ shown in Fig. \ref{Fc5}.
}
\end{figure}

\section{Absence of accidental degeneracy}\label{accidental}

The energy spectra of {\em any} central potential exhibit
the $(2\ell+1)$-fold ''essential degeneracy'' as the
energy levels $E_{k,\ell}$ do not depend on the magnetic
quantum number $m$. It has been proven long ago
that the only potentials that show {\em accidental}
degeneracy are the Coulomb and the harmonic oscillator one, respectively.\cite{Bertrand}
This is a consequence of the existence of a conserved quantity which does
not commute with any member of a complete system of commuting operators
of the problem.\cite{LandauIII}
In the Coulomb case this is the well-known
Runge-Lenz vector.\cite{Runge, Lenz} The conserved quantity for the
harmonic oscillator problem we shall discuss below.
We conclude that in our potential which is neither of both
accidental degeneracy is absent. Nevertheless we shall study
this problem in more detail here.

\subsection{Trajectory of a classical particle}\label{degeneracy}

\begin{figure}[htb]
\includegraphics[height=8cm,width=8cm]{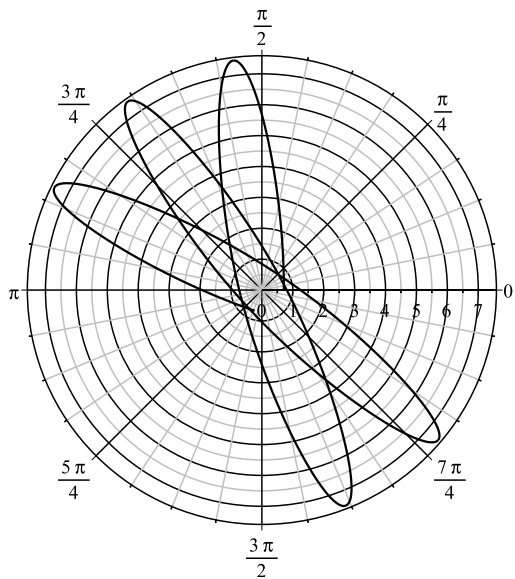}
\caption{\label{Fc5}
Trajectory $r(\Theta)$ of a classical particle with mass $\mu$ and energy 
$E=0.86\,V_0$ moving in the effective potential shown in Fig. 4.
The trajectory starts at an inner turning point and an angle $\Theta=0$.
After having covered five periods it reaches an inner turning point at an angle 
$\Theta\approx 11\pi/8$.
}
\end{figure}

We recall the trajectories  of a classical particle in
the harmonic oscillator  as well as in the Coulomb potential are
closed, the latter for negative energies only. Following the textbook \onlinecite{Goldstein}
we calculate the trajectory of a classical particle of mass $\mu$,
energy $E$ and angular momentum $J$ moving in
the effective potential 
\begin{equation}
V_{\rm eff}(r) =\frac{J^2}{2\mu r^2}+V_0 \,v(r)
\end{equation}
with an energy $E=0.86\,V_0$ periodically
between both turning points
as displayed in Fig. \ref{Fc4}
while
five periods of the trajectory $r(\Theta)$ are 
shown in Fig. \ref{Fc5} where
the dimensionless radius 
$\rho = \alpha_{\rm cl}\, r$ with
\begin{equation}
\alpha_{\rm cl}=\left(\frac{\mu V_0}{J^2}\right)^{1/2}
\end{equation}
was used.

The potential determined numerically from the spectrum (\ref{SinglePartSpect})
and shown in Fig. 2 is
denoted by $v(\rho)$. It is evident that the orbit of the particle 
does not close but precesses around the force center thus indicating the absence
of accidental degeneracy.

\subsection{Energy spectrum}

The most direct way to check for degeneracy is simply to calculate
the energies $E_{k,\ell}$ for the potential under
consideration with radial and azimuthal quantum numbers $k$  and $\ell$,
respectively. If two ore more of the energies with different indices are equal
degeneracy is present otherwise not.
\begin{figure}[htb]
\includegraphics[height=8cm,width=8cm]{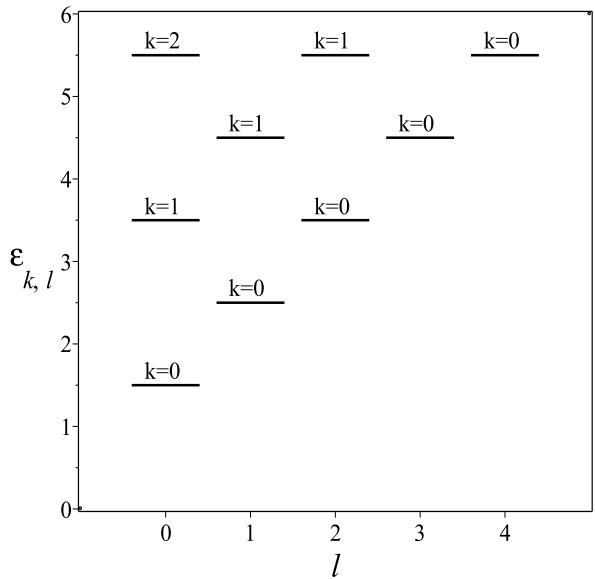}
\caption{\label{Fc6}
Lowest scaled energies of the three-dimensional harmonic oscillator.
The scheme of levels $E_n=\hbar \omega(n+3/2)$ shows degeneracy as the
principal quantum number $n=2k+\ell$ depends on both,
the radial quantum number $k$ and the azimuthal quantum number $\ell$,
respectively. For example the level $n=2$ is doubly degenerate
for the quantum numbers $k=1, \ell=0$ and $k=0, \ell=2$, respectively.
}
\end{figure}
\begin{figure}[htb]
\includegraphics[height=8cm,width=8cm]{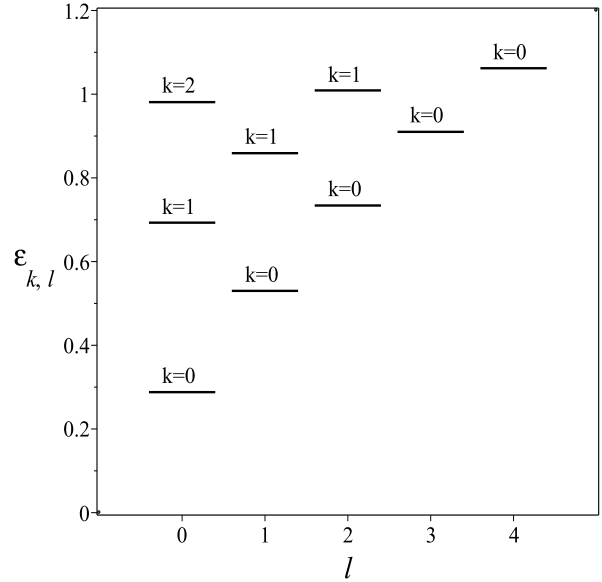}
\caption{\label{Fc7}
Scaled energies of a particle with mass $\mu$ moving in a
three-dimensional potential leading to a spectrum with $s$ state
part (\ref{Etilde}) and scaling parameter
$K=2$ of Eq. (\ref{Lprime}). Every energy level is characterized by
{\em two} quantum numbers $k$ and $\ell$, respectively.
No principal quantum number can be identified and evidently
no accidental degeneracy takes place.
}
\end{figure}

Before we turn to our potential $v(\rho)$ we recall the situation for the
three-dimensional harmonic oscillator where the lowest energy levels are
displayed in Fig. \ref{Fc6}. 

The energies $E_{k,\ell}$ depend on a combination of both indices
$k$ and $\ell$ namely on the principal quantum number $n=2k+\ell$ 
leading to degeneracy of the levels $E_n=\hbar \omega (n+3/2)$
as can be checked from levels with $n=2,3,4$ of the figure.\cite{Cohen}
If the x- and y-axis are oriented along the symmetry axes of the elliptic orbit 
of the oscillator
then it can be shown that the additional
integral of motion reduces to the scalar function $E_x - E_y$, the difference between the
energies of the projections of the motion onto the x- and the y-axis, respectively.\cite{Khriplovich} 

With the help of the potential $v(\rho)$ we solved the radial equation (\ref{RadialEq})
numerically. The lowest energy levels $E_{k,\ell}$ are displayed in Fig. \ref{Fc7}.
At first sight the scheme of the energies resembles that of the harmonic oscillator potential.
But looking more closely we observe
that the levels which in the harmonic oscillator scheme of Fig. \ref{Fc6}
were degenerate with each other now differ slightly. We
conjecture that the higher energy levels behave similarly and 
no accidental degeneracy is present. We emphasize
once more that the $(2\ell+1)$-fold {\em essential} degeneracy
with respect to the magnetic quantum number $m$ is 
caused by the central potential $v(\rho)$.

\end{document}